\documentclass[a4, 11pt]{article}
\usepackage{amsmath} %Never write a paper without using amsmath for its many new commands
\usepackage{amssymb} %Some extra symbols
\usepackage{geometry} % to change the page dimensions
\usepackage{graphicx} % support the \includegraphics command and options
\usepackage{epsfig}
\usepackage{booktabs} % for much better looking tables
\usepackage{array} % for better arrays (eg matrices) in maths
\usepackage{paralist} % very flexible & customisable lists (eg. enumerate/itemize, etc.)
\usepackage{verbatim} % adds environment for commenting out blocks of text & for better verbatim
\usepackage{subfig} % make it possible to include more than one captioned figure/table in a single float
\usepackage[dutch]{babel}
\usepackage{fancyhdr} % This should be set AFTER setting up the page geometry
\usepackage{sectsty}
\allsectionsfont{\sffamily\mdseries\upshape} % (See the fntguide.pdf for font help)
\usepackage[titles,subfigure]{tocloft} % Alter the style of the Table of Contents

\title{Methodology for Simulation-based Comparison of Algorithms for Distributed Mutual Exclusion}
\author{Filip De Turck, Ghent University - imec, Belgium}   
\date{}
\begin{document}
\maketitle
\renewcommand{\abstractname}{Abstract}
\renewcommand{\refname}{References}
\begin{abstract} 
In this paper, we show how different types of distributed mutual algorithms can be compared in terms of performance through simulations. A simulation-based approach is presented, together with an overview of the relevant evaluation metrics and approach for statistical processing of the results. The presented simulations can be used to learn master students of a course on distributed software the basics of algorithms for distributed mutual exclusion, together with how to properly conduct a detailed comparison study. Finally, a related work section is provided with relevant use cases where distributed mutual exclusion algorithm can be beneficial.
\end{abstract}

\section{Problem statement}
Consider a common resource (e.g. an actuator, a camera, etc), where only one out of $N$ processes may be simultaneously using this resource. This requirement implies that these $N$ processes coordinate their access and that access is only granted to one and only one process at any given time. The block of code, where exclusive access is to be guaranteed, is called a critical section. This problem is known as the mutual exclusion problem, which, in the case of a standalone application, is solved using operating system primitives (mutex, semaphores, etc.). The problem we now tackle is distributed mutual exclusion, where the competing processes only communicate through message passing. The used failure model assumes: (i)~a reliable channel between communicating processes, (ii)~processes do not crash, (iii)~processes eventually leave the critical section (processes are well behaving).

\section{Evaluation metrics}
Next to the number of exchanged messages, there are two important metrics to characterize the performance of algorithms for distributed mutual exclusion:
\begin{itemize}
\item Client delay: denotes how much time does it take for a client to enter/leave the critical section, in case the critical section is not used by another client, where the system is assumed to be unloaded. The client delay is most influenced by the network delay (the one-way network delay is denoted as $\delta$) and the processing delay. 

\item Synchronization delay: this metric gives an indication how much time the algorithm wastes between subsequent accesses to the critical section. The synchronization delay is the time (typically expressed as a function of $\delta$) needed between one client leaving the critical section, and the next client accessing the critical section. Here, we suppose that the system is heavily loaded (to allow fair comparison). Obviously, this synchronization delay puts an upper bound on the number of clients that can be served per unit of time.
\end{itemize}

\section{Three considered algorithms}
Several algorithms exist for distributed mutual exclusion, three of them are detailed below.

\subsection{Central Server Algorithm}
In the Central Server algorithm, one process takes the role of coordinator, receiving requests to access the critical section from all other processes, and granting access. Often, centralized approaches are simple in terms of algorithmic complexity, but exhibit poor scaling behaviour.

\subsection{Ring Algorithm}
In a ring based approach, the communicating processes are organized in a logical ring, and only communicate with one neighbour (Figure 3): in case the ring consists of N processes ${p_0, . . . , p_{N-1}}$, process $p_i$ has a single, unidirectional communication channel to process $p_{(i+1)mod N}$. The token, granting access now travels along the ring. There is only one message type exchanged, i.e. the message Token, so we only have to specify what happens when this message is received. Also note that all processes implement the same logic (there is no central server), making this algorithm suitable to dynamic adaptations.

\subsection{Raymond's Algorithm}
This algorithm is a token-passing algorithm for tree-based networks. It is assumed that processes are connected through a tree topology, and a token is passed between processes. A node that holds the token, serves a root of the tree, and child nodes maintain a pointer to the root node. Each node has two variables: (i)~a parent pointer, this pointer is empty in the root node and non-empty in the child nodes, and (ii) a local queue Q to store pending requests. 
A node that wants access to the critical, sends a request towards the root via its parent. In each visited node, pointers to the root node are followed. Only the first request in Q is forwarded to the parent node. When a token moves from one process to another, the following two steps are taken: (i)~the root changes, by swapping the parent variables between the processes, and (ii)~the node with the token becomes the new root.

\section{Simulation approach}
Assume that the communication delay $\delta$ always follows a Gaussian distribution with a mean of 30 milliseconds and standard deviation of 5 milliseconds.
Consider the following three algorithms for distributed mutual exclusion:
\begin{enumerate}

\item the central server algorithm, where the processing time in the central server equals $40 \times e^{N/10}$ milliseconds (for both request- and leave-messages), where $N$ denotes the number of clients involved and $e^x$ denotes the exponential function. 

\item the ring-based algorithm with $N$ different nodes.  The processing delay in each node (to receive and send the token, or receive and keep the token) follows a gaussian distribution with a mean of 15 milliseconds and standard deviation of 2 milliseconds.

\item the Raymond's algorithm for distributed mutual exclusion, with N different nodes. The processing delay in each node (to receive and send the token, or receive and keep the token) follows a gaussian distribution with a mean of 15 milliseconds and standard deviation of 2 milliseconds. Each node has a pointer to the parent node (value=NULL for the root node), and two pointers to the left child node and the right child node, respectively. 

\end{enumerate}

Assume $N$ equals 100, implement the three algorithms above in the C++ programming language and keep track of the client delay and the synchronization delay for 50 different requests. Determine the average client delay and synchronization delay, together with the standard deviation for the three algorithms.\\

\section{HPC infrastructure}
All examples are supposed to be compiled and executed using the UGent HPC (High Performance Computing) infrastructure. Once connected to the HPC infrastructure, students are logged on to one of the interface nodes. These nodes can be used to compile software and submit jobs to the different clusters, but running software on these interface nodes is generally considered bad practice. For this assignment however, it is fine to run lightweight examples on the interface node. Hence, you may simply execute the commands in the assignments on the interface nodes. The project should be compiled and run on a interface nodes. In order to compile all source files, please use the provided compile.sh script (that relies on CMake internally to generate a Makefile).

\section{Discussion}
The comparison study of the three algorithms for distributed mutual exclusion gives insight in the following elements:

\begin{enumerate}

\item the internal algorithmic details of the algorithms (centralized, ring-based and tree-based),

\ item the typical simulation-based modelling of the communication delays and processing delays,

\item the evaluation setup and statistical processing of the obtained numerical results, and

\item typical numerical values for the two important evaluation metrics: client delay and synchronization delay.

\end{enumerate}

An important guideline is to clearly guide the simulation study, and emphasize the difference with a typical emulation study.\\
Distributed mutual exclusion algorithms are beneficial in several application domains. Important use cases where distributed mutual exclusion algorithms can be applied are: mobile augmented reality applications~\cite{JSSVerbelen}, hierarchical network management systems~\cite{hierMoens, hiermgmtsystem}, scalable and adaptive video delivery~\cite{scalablevideolaga, 6dofadaptive, reviewMariaQoE}, efficient resource management for virtual desktop cloud computing~\cite{SupercomputingDeboosere}, smart city applications~\cite{CNSM2017Santos, NOMS2018Santos},  replica placement in ring based content delivery networks~\cite{ComComWauters}, and softwarized networks~\cite{vnfp, tnsmmoens}.

\section{GitHub repository}
\texttt{https://github.ugent.be/fdeturck/PDS/tree/main/HWA3}

\end{document}